\documentclass[12pt,aps,preprint,nofootinbib,superscriptaddress,nobalancelastpage]{revtex4}
\pdfoutput=1

\usepackage{mathrsfs}
\usepackage{amsmath,amsthm,amssymb}
\usepackage{color}
\usepackage{graphicx}
\usepackage{verbatim}

\usepackage{hyperref}
\hypersetup{colorlinks,bookmarksopen,bookmarksnumbered,citecolor=blue,
linkcolor=blue,pdfstartview=FitH,urlcolor=blue}

\newcommand{\be}{\begin{equation}}
\newcommand{\ee}{\end{equation}}
\newcommand{\bea}{\begin{eqnarray}}
\newcommand{\eea}{\end{eqnarray}}

\newcommand{\fig}{Fig.}
\newcommand{\Sec}{Sec.}
\newcommand{\Ref}{Ref.}
\newcommand{\Refs}{Refs.}
\newcommand{\eq}{Eq.}

\newcommand{\ie}{\emph{i.e.}}

\newcommand{\NOvA}{NO$\nu$A}

\newcommand{\ESS}{ESS$\nu$SB}

\begin{document}
\preprint{FTUAM-14-23, IFT-UAM/CSIC-14-060, NORDITA-2014-80}

\title{Reassessing the sensitivity to leptonic CP violation}

\author{Mattias Blennow}
\email{emb@kth.se}
\affiliation{Department of Theoretical Physics, School of Engineering Sciences, KTH Royal Institute of Technology, AlbaNova University Center, 106 91 Stockholm, Sweden}
\author{Pilar Coloma}
\email{pcoloma@vt.edu}
\affiliation{Center for Neutrino Physics, Physics Department, Virginia Tech, 850 West Campus Dr, Blacksburg, VA 24061, USA}
\author{Enrique Fernandez-Martinez}
\email{enrique.fernandez-martinez@uam.es}
\affiliation{Departamento de F\'isica Te\'orica, Universidad Aut\'onoma de Madrid, Cantoblanco E-28049 Madrid, Spain}
\affiliation{Instituto de F\'isica Te\'orica UAM/CSIC,
 Calle Nicol\'as Cabrera 13-15, Cantoblanco E-28049 Madrid, Spain}

\begin{abstract}
We address the validity of the usual procedure to determine the sensitivity of neutrino oscillation experiments to CP violation. An explicit calibration of the test statistic is performed through Monte Carlo simulations for several experimental setups. We find that significant deviations from a $\chi^2$ distribution with one degree of freedom occur for experimental setups with low sensitivity to $\delta$. In particular, when the allowed region to which $\delta$ is constrained at a given confidence level is comparable to the whole allowed range, the cyclic nature of the variable manifests and the premises of Wilk's theorem are violated. This leads to values of the test statistic significantly lower than a $\chi^2$ distribution at that confidence level. On the other hand, for facilities which can place better constraints on $\delta$ the cyclic nature of the variable is hidden and, as the potential of the facility improves, the values of the test statistics first become slightly higher than and then approach asymptotically a $\chi^2$ distribution. The role of sign degeneracies is also discussed.
\end{abstract}

\maketitle

\section{Introduction}
\label{sec:intro}

Mixing in the lepton sector of the Standard Model is described by the unitary Pontecorvo--Maki--Nakagawa--Sakata~(PMNS) matrix~\cite{Pontecorvo:1957cp,Pontecorvo:1957qd,Maki:1960ut,Maki:1962mu,Pontecorvo:1967fh}. In the standard three family scenario, it can be parametrized by three mixing angles ($\theta_{12}$, $\theta_{23}$, and $\theta_{13}$), a Dirac CP violating phase ($\delta$) and, if neutrinos are Majorana particles, two additional Majorana CP phases. Neutrino oscillations are sensitive to the values of all mixing angles and the CP violating phase $\delta$, together with the two independent mass squared differences ($\Delta m_{21}^2$ and $\Delta m_{31}^2$, defined as $\Delta m_{ij}^2 \equiv m_i^2-m_j^2$). Unlike for the CKM matrix, the mixing angles of the PMNS matrix have been experimentally found to be large, with $\theta_{13} \simeq 9^\circ$ being the smallest mixing angle~\cite{An:2012eh,Ahn:2012nd,Abe:2012tg,Adamson:2011qu,Abe:2011sj} and the other two angles being much larger ($\theta_{12} \sim 33^\circ$ and $\theta_{23}\sim 45^\circ$~\cite{GonzalezGarcia:2012sz}). Thus, the Jarlskog invariant $J = \cos \theta_{13} \sin 2 \theta_{13} \sin 2 \theta_{12}\sin 2 \theta_{23} \sin\delta$ can be potentially as large as $\sim 0.035$ for maximally CP violating values of $\delta$, three orders of magnitude higher than the currently measured value for its counterpart in the quark sector: $J_{\rm CKM} = (2.96^{+0.20}_{-0.16})\cdot 10^{-5}$~\cite{Beringer:1900zz}. Since it has been shown that, within the context of Standard Model Electroweak Baryogenesis, $J_{\rm CKM}$ is not large enough to account for the observed Baryon asymmetry of the Universe~\cite{Gavela:1993ts,Gavela:1994dt}, the discovery of an additional source of CP violation (such as $\delta$ in the PMNS matrix) could open new possibilities for alternative generation mechanisms.

Given the current knowledge on the neutrino oscillation parameters, the focus for the next generation of neutrino oscillation experiments will be to determine the neutrino mass ordering, the CP violating phase $\delta$, and the octant of the mixing angle $\theta_{23}$. For the determination of the mass ordering, there has been a lively discussion in the literature on whether or not the common way of assessing the sensitivity of a future experiment is applicable~\cite{Qian:2012zn,Capozzi:2013psa,Blennow:2013oma,Vitells:2013uza}. As shown in \Ref~\cite{Blennow:2013oma}, the common approach does give a reasonable approximation of the median sensitivity, although not for the reasons typically used in the argumentation for it. The small correction was mainly based on the one-sided nature of the hypothesis test and only to a minor degree on the non-gaussianity of the statistical distributions. Similarly, deviations from a $\chi^2$ distribution of the test statistic should be tested for in the search for $\delta$. Indeed, one of the requirements for the validity of the common approach when making sensitivity analyses is that the change in number of events forms a linear space upon variations of $\delta$. However, this requirement will necessarily be violated at some level since $\delta$ is periodic and a change by $2 \pi$ will leave the number of events unchanged. In addition, there is no guarantee that the predicted data without statistical fluctuations, as used in the common approach, will be representative. In the present work, these assumptions are tested by explicit Monte Carlo simulations in order to find out exactly how much the sensitivity analyses are affected by these assumptions. To do so, we start from the basic frequentist definitions and apply the Feldman--Cousins approach~\cite{Feldman:1997qc} in order to determine the sensitivity of several experimental setups.

\section{Statistical approach}
\label{sec:statistics}

The most common way of quantifying the experimental sensitivity to leptonic CP violation is to quote the confidence level at which the CP conserving values of $\delta$ (0 and $\pi$) can be rejected. In the literature, this is typically computed by constructing the test statistic
\begin{equation}
\label{eq:teststatistic}
 S = \min_{\delta = 0, \pi} \chi^2 - \min_{\rm global} \chi^2,
\end{equation}
where $\chi^2 = - 2\log \mathcal L$, and $\mathcal L$ is the likelihood of observing the data given a particular set of oscillation parameters. 
It is worth noting that this involves minimizing over all nuisance parameters, which may be subject to external constraints (we will discuss how such external constraints are handled in \Sec~\ref{sec:external}).
It is then assumed that $S$ is $\chi^2$-distributed with one degree of freedom, based on the implications of Wilks' theorem~\cite{Wilks:1938dza}. In addition, the Asimov data set\footnote{So named after the \emph{Franchise} short story by Isaac Asimov, where an entire electorate was replaced by one single representative.}~\cite{Cowan:2010js}, \ie, the event rates without statistical fluctuations, is assumed to be representative for the experimental outcome and is thus used to estimate the expected confidence level~(CL) at which the CP conservation hypothesis would be rejected. 
However, as for the case of the neutrino mass ordering~\cite{Qian:2012zn,Blennow:2013oma}, it is not clear to what degree the assumptions underlying Wilks' theorem are violated when testing CP conservation in this fashion, resulting in a need to explicitly test this framework.

The procedure to test the CP conservation hypothesis at a given CL can be summarized as follows: First, the distribution of the test statistic $S$ is found by simulating a large number of realizations of the experiments based on the predicted event rates under the assumption of CP conservation (\ie, for $\delta = 0,\pi$). This is done for a given set of values for the other oscillation parameters, which we assume to be the true values. The value of $S$ is then computed for each realization, which provides the distribution of $S$. CP conservation will be rejected at CL\ $x$ if the measured value of $S$ is among the $1-x$ fraction of largest values in the distribution. This automatically defines a critical value, $S_c(x)$, such that CP conservation is rejected at CL\ $x$ if $S > S_c(x)$. By construction, $S_c(x)$ is the inverse of the cumulative distribution function~(CDF) of $S$ under the CP conserving hypothesis. 

The above construction is only concerned with the test of CP conservation for a given data set, \ie, once the experiment has already taken data. The expected performance of future facilities will depend on the true value of $\delta$. Thus, in a frequentist approach the performance of the facility must be quantified for each value of $\delta$ separately. In addition, due to statistical fluctuations, different realizations of a given facility will lead to a different significance at which CP conservation can be rejected. Therefore, the convention is to define the expected sensitivity of a given experiment as the CL\ obtained for the median of the distribution, and is typically shown as a function of the value of $\delta$ itself. This is usually referred to as the median sensitivity and it will not necessarily coincide with the significance computed with the Asimov data set. 

A calibration of the $\chi^2$ for CP violation, following the procedure described above, was performed in \Ref~\cite{Schwetz:2006md}. The T2HK experiment~\cite{Abe:2011ts} was used as an example, and the critical values were found to be significantly smaller than for the $\chi^2$ distribution with 1 degree of freedom in the region $\sin^22\theta_{13} \gtrsim 10^{-2}$. However, the study in \Ref~\cite{Schwetz:2006md} was restricted to values of $\sin^22\theta_{13}\lesssim 5\times 10^{-2}$, and was done using a different test statistic than the one considered in the present work.

\subsection{External constraints}
\label{sec:external}

The common approach when dealing with external constraints on the nuisance parameters (such as previous determinations of oscillation parameters or prior constraints for the systematic uncertainties) is to include an additional term in the $\chi^2$ of the form
\begin{equation}
 \chi^2 = \chi^2_0(\xi) + \frac{(\xi - \bar\xi)^2}{\sigma_\xi^2},
 \label{eq:chi2}
\end{equation}
where $\xi$ is the nuisance parameter, $\chi_0^2(\xi)$ is the $\chi^2$ provided by the experiment itself for a given value of $\xi$, and $\sigma_\xi$ is the error in the determination of $\xi$ which has a measured central value of $\bar\xi$. In order to calibrate the $\chi^2$ to do a proper hypothesis test, statistical fluctuations must also be considered for the experiment determining $\bar\xi$. From this follows that, when calibrating the critical value $S_c$ for an assumed true value $\xi_{\rm true}$, the values of $\bar\xi$ should be chosen according to a normal distribution with mean $\xi_{\rm true}$ and standard deviation $\sigma_\xi$.\footnote{Note that this is \emph{not} completely equivalent to picking $\xi$ from a random distribution. The test statistic must still be separately calibrated for each simple hypothesis.} The final $\chi^2$ is obtained after marginalizing over the nuisance parameters $\xi$ in \eq~\ref{eq:chi2}. The test statistic is then defined through \eq~(\ref{eq:teststatistic}) and the critical value is computed accordingly.

When computing the median outcome expected for a CP violating value of $\delta$, however, the outcome of any external \emph{past} experiment should be taken into account. This implies that $\bar\xi$ should be fixed to the result from the external experiment. By doing this, one would obtain the probability of reaching a given CL using the already known outcome for the external constraints. This is the procedure that has been followed in this paper. On the other hand, if external constraints from a hypothetical \emph{future} experiment were to be implemented instead, the outcome of such an experiment would be unknown and should therefore still be chosen according to the expected distribution of possible outcomes. 

In our simulations, we have followed the prescription described above for existing constraints on the neutrino oscillation parameters as well as for including constraints on the systematic errors. The calibration of the test statistic has been performed for the best fit values of the parameters. Given our present knowledge, we do not expect the calibration of the test statistic for CP violation to be crucially dependent on this choice with the notable exception of $\theta_{23}$, for which strong correlations with $\delta$ exist~\cite{Coloma:2014kca} and preliminary explorations show that this can strongly reflect in the test statistic calibration~\cite{Gonzalez-Garcia:2014bfa}. Such correlations call for a calibration of the test statistic as a function of both $\delta$ and $\theta_{23}$ in order to properly asses the significance of a signal in both variables simultaneously. Nevertheless, such a study is computationally very expensive and therefore beyond the scope of this work. 

\section{Simulation details}

We simulate the long-baseline experiments LBNE~\cite{Adams:2013qkq}, T2HK~\cite{Abe:2011ts}, \ESS~\cite{Baussan:2013zcy}, and \NOvA ~\cite{Ayres:2004js} using the GLoBES software~\cite{Huber:2004ka,Huber:2007ji} to find their respective Asimov data sets as functions of the parameter values. The simulation details for the LBNE, T2HK, \ESS, and \NOvA\ experiments were implemented as in \Refs~\cite{Blennow:2013oma}, \cite{Coloma:2012ji}, \cite{Baussan:2013zcy}, and \cite{Blennow:2013oma}, respectively\footnote{The only modification is that in the present work the beam power of the LBNE experiment has been increased to 1.2~MW, and the detector mass has been fixed to 34~kt.}. For the \ESS, we use the configuration with a 540~km baseline and 2.5~GeV protons. Once the Asimov data sets were computed, realizations of the experiments were constructed by applying Poisson statistics individually to the number of events in each bin based on the expected rates. This was implemented using the MonteCUBES software~\cite{Blennow:2009pk}. The value of $S$ was then computed for each realization in order to find the distribution of $S$ under the CP conserving and CP violating hypotheses. In all cases, 5~\% (10~\%) systematic errors were used for the signal (background) event rates. These are bin-to-bin correlated, but uncorrelated between signal and backgrounds and between different oscillation channels. The true values of the oscillation parameters have been set according to the best fits in \Ref~\cite{GonzalezGarcia:2012sz}, with the exception of $\theta_{23}$ which has been set to $45^\circ$. Marginalization is performed on $\theta_{12}$, $\sin^22\theta_{13}$, $\sin^22\theta_{23}$, $\Delta m^2_{21}$ and $\Delta m^2_{31}$, using gaussian priors in agreement with present experimental uncertainties from \Ref~\cite{GonzalezGarcia:2012sz}. Sign degeneracies are fully taken into account during minimization. 

In order to compute the critical values of $S$, we simulate $10^5$ realizations for $\delta=0$ and $\pi$ and for both neutrino mass orderings. This is sufficient to reliably find the value of $S_c$ up to $\sim 3.5\sigma$ level, corresponding to a CL of $\sim 99.95~\%$.
On the other hand, for the computation of the expected outcomes for CP violating values of $\delta$ we are only interested in obtaining the value of $S$ for the median of the distributions. Since this is much less sensitive to statistical fluctuations than the sampling of the tails, only $10^3$ realizations of the experiments are simulated in order to obtain the median of $S$ for these values.

\section{Results}

In this section we present our simulation results and we discuss the general behaviour of the test statistic and the final sensitivities for CP violation. In \Sec~\ref{sec:dist} we show the distribution of the test statistic defined in \eq~\ref{eq:teststatistic} for all facilities under consideration, as obtained from the Monte Carlo simulations. Then, in \Sec~\ref{sec:discuss} we discuss the dependence of the distributions with several factors, such as the statistics of a given experiment and/or the presence of sign degeneracies. Finally, in \Sec~\ref{sec:sens} we show the resulting sensitivities obtained from Monte Carlo and we compare to the usual values reported in the literature (which are obtained under the assumption that Wilks' theorem is valid).

\subsection{Distribution of the test statistic for the null hypothesis}
\label{sec:dist}

\begin{figure}
 \begin{center}
  \includegraphics[width=0.7\textwidth]{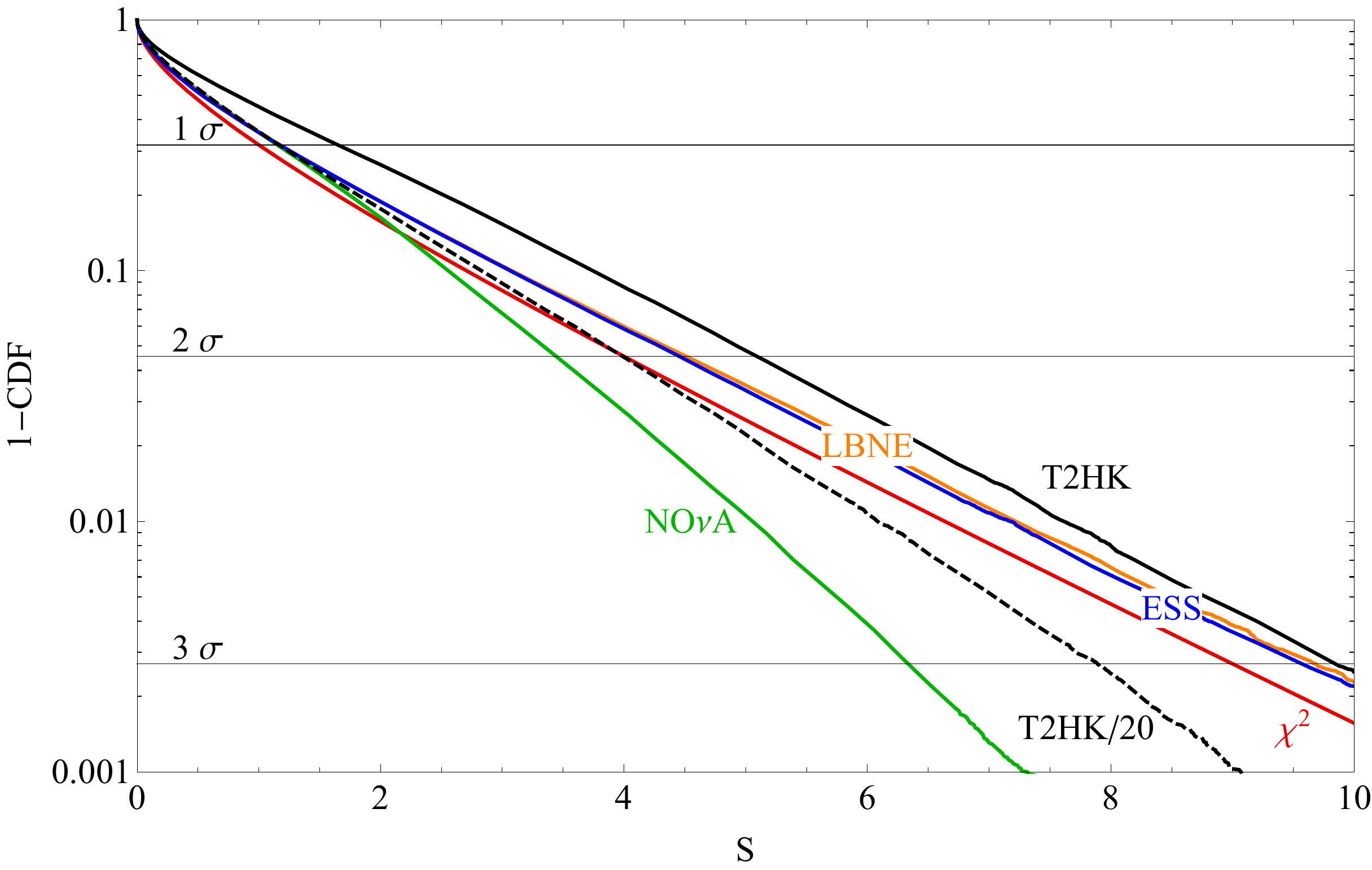}
  \caption{Distribution of the test statistic $S$ for the different simulated experiments. We show the value of $1-{\rm CDF}(S)$, where CDF is the cumulative distribution function. The values of $1-{\rm CDF}$ corresponding to $1\sigma$, $2\sigma$, and $3\sigma$ are shown as horizontal lines. For comparison, the red line shows the result for a $\chi^2$ distribution with one degree of freedom. \label{fig:cdfs}}
 \end{center}
\end{figure}

Figure~\ref{fig:cdfs} shows our results for the distribution of the test statistic $S$ for the experimental setups considered in this work. In order to show the dependence with the statistics of the experiment, results are also shown for T2HK with a factor 20 reduced statistics.  Such a setup would also be similar to T2K at the end of its planned running time, although with a somewhat extended run and equal running times in the neutrino and antineutrino modes.
As can be seen from the figure, the CDFs are generally close to a $\chi^2$ distribution with one degree of freedom for almost all experiments under consideration. The notable exceptions to this rule are the NO$\nu$A setup and T2HK with reduced statistics, for which large deviations are observed and the critical values corresponding to a given CL are considerably smaller than the values obtained under the assumption of a $\chi^2$-distributed test statistic. 
This can be understood as follows. One of the requirements for the applicability of Wilks' theorem is that, when varying the parameter that is being tested for (in this case $\delta$), the subsequent change in the observables used to determine it should constitute a linear space. This requirement will obviously be violated at some level for $\delta$, since a change by $2 \pi$ will render no change in the number of events. Nevertheless, if a given facility is able to constrain the value of $\delta$ very precisely, the linearity condition is expected to be approximately satisfied for the statistical fluctuations on the number of events. Therefore, one could naively expect that the deviations from a $\chi^2$ distribution would be more manifest for experiments with the poorest sensitivity to $\delta$, where the violation of the requirements for Wilks' theorem is more apparent. This will be discussed in more detail in \Sec~\ref{sec:discuss}. 

The other relatively significant deviation from a $\chi^2$ distribution can be seen for T2HK. We have checked that this extra deviation with respect to the better agreement showed by LBNE and the ESS can be attributed to the sign degeneracies, that play a very important role in the determination of $\delta$ from T2HK. A parametric degeneracy will again lead to situations in which the change in the number of events does not span a linear space, implying the non-applicability of Wilks' theorem. When the mass hierarchy is assumed to be known, the distribution obtained for T2HK lies on top of those obtained for LBNE and the ESS as expected.

\subsection{Discussion}
\label{sec:discuss}

In this section, we explicitly test the interpretation of \fig~\ref{fig:cdfs} presented above with detailed simulations and geometric arguments based on the cyclicity of $\delta$ and thus on the expected violations of Wilk's theorem.
\begin{figure}
 \begin{center}
 \begin{tabular}{cc}
  \includegraphics[width=0.45\textwidth]{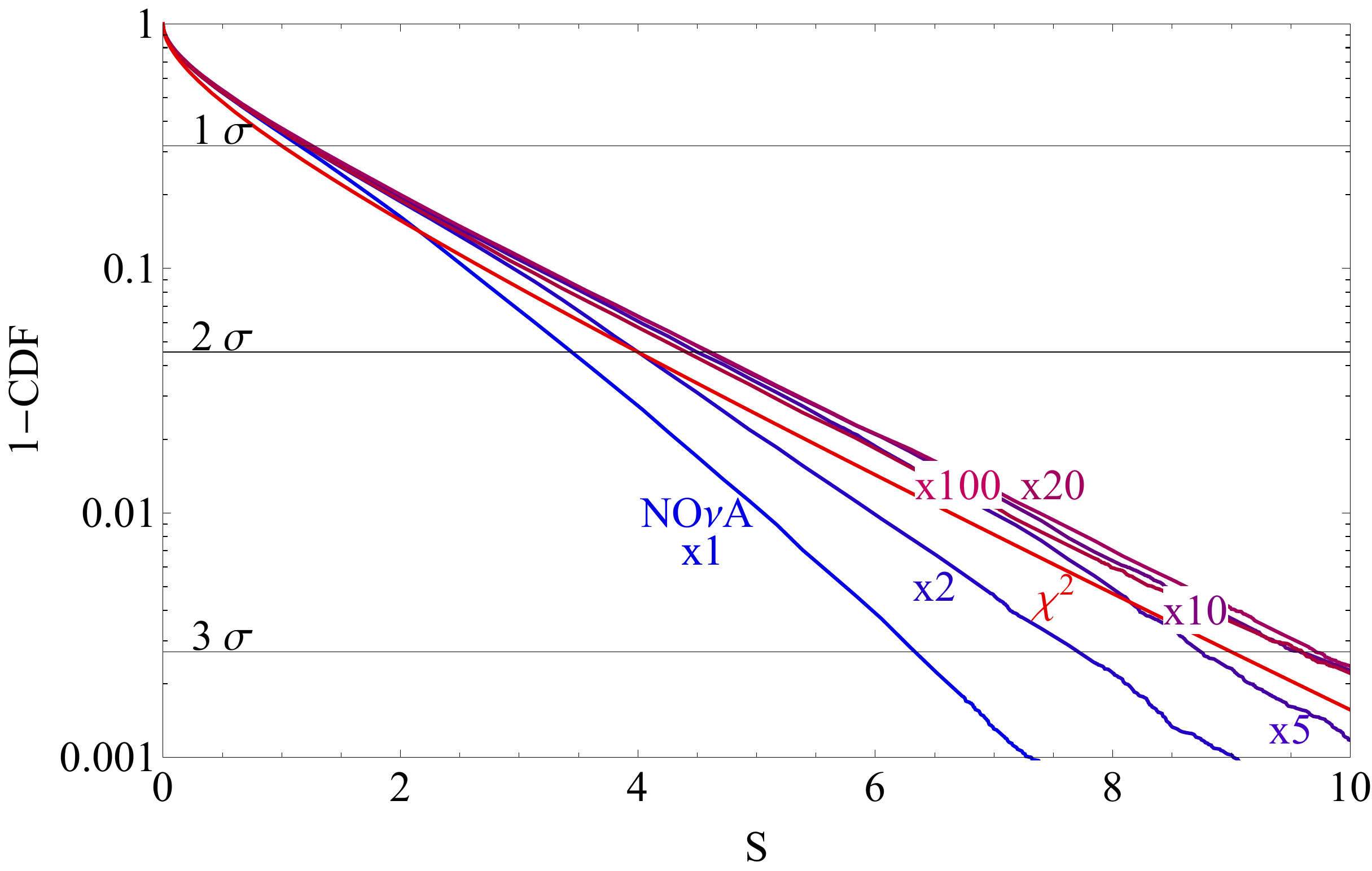} &
  \raisebox{-.05\height}{\includegraphics[width=0.46\textwidth]{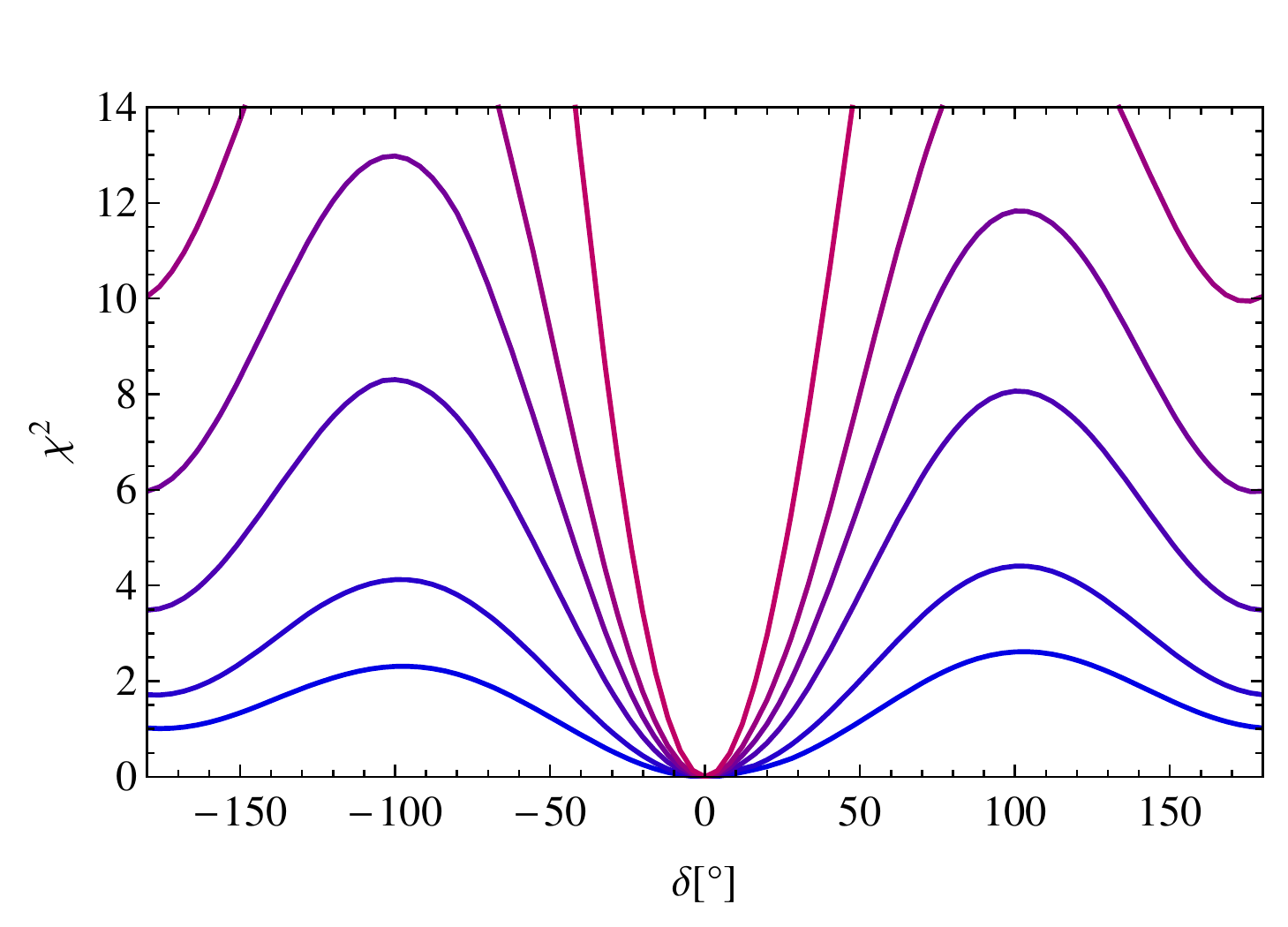}}
  \end{tabular}
  \caption{Left panel: distribution of the test statistic $S$ for the NO$\nu$A setup increasing its nominal exposure by factors of 1, 2, 5, 10 and 100. We show the value of $1-{\rm CDF}(S)$, where CDF is the cumulative distribution function. The values of $1-{\rm CDF}$ corresponding to $1\sigma$, $2\sigma$, and $3\sigma$ are shown as horizontal lines. For comparison, the red line shows the result for a $\chi^2$ distribution with one degree of freedom. Right panel: $\chi^2$ profile for the rescaled NO$\nu$A setups as a function of $\delta$ for a true value of $\delta=0$, obtained in absence of statistical fluctuations. As in the left panel, the different lines have been obtained after increasing the nominal exposure by factors of 1, 2, 5, 10, 20 and 100. \label{fig:novacheck}}
 \end{center}
\end{figure}
In the left panel of \fig~\ref{fig:novacheck} we take NO$\nu$A as a test setup and increase its statistics by factors of 1, 2, 5, 10, 20 and 100 in order to improve the corresponding determination of $\delta$ and thus to quantify how precise this determination needs to be so as to recover a distribution close to a $\chi^2$. As can be seen in the left panel, upon increasing the statistics assumed for NO$\nu$A, the deviation from a $\chi^2$ distribution becomes milder and happens at higher and higher CL. Indeed, the different curves show a change of trend developing a sharper decrease after a certain confidence level, which increases with statistics. For an increase in statistics of one order of magnitude this deviation is no longer seen for confidence levels below $3 \sigma$, which are the ones we are able to probe with the number of experimental realizations simulated in this work. Indeed, for increases of the statistics by factors of 10 and 20 the obtained distribution is rather consistently \emph{above} a $\chi^2$ distribution, while an increase of statistics by a factor 100 brings it down and therefore closer to the $\chi^2$.

In the right panel of \fig~\ref{fig:novacheck}, we show the correlation between the observed deviations from the $\chi^2$ distribution for the various exposures with the precision with which they would be able to reconstruct $\delta$, for a true value of $\delta = 0$ and in absence of statistical fluctuations. As expected from the cyclic nature of $\delta$, a degeneracy between $\delta=0$ and $\delta=180^\circ$ tends to take place in all cases. It can be seen that the height of the barrier separating the two minima in the right panel seems roughly correlated with the value of the test statistic for which a change of trend is observed in the slope of the curves in the left panel. The change in slope is not related to the degeneracy between $\delta =0,\pi$ itself, but rather to the fact that the reconstructed interval in $\delta$ at such CL would become of the same order of the whole allowed range for this variable. Therefore, its cyclicity and hence the violation of the requisites of Wilk's theorem become manifest at the CL given by the height of the barrier separating the two minima.

\begin{figure}
 \begin{center}
  \includegraphics[width=0.35\textwidth]{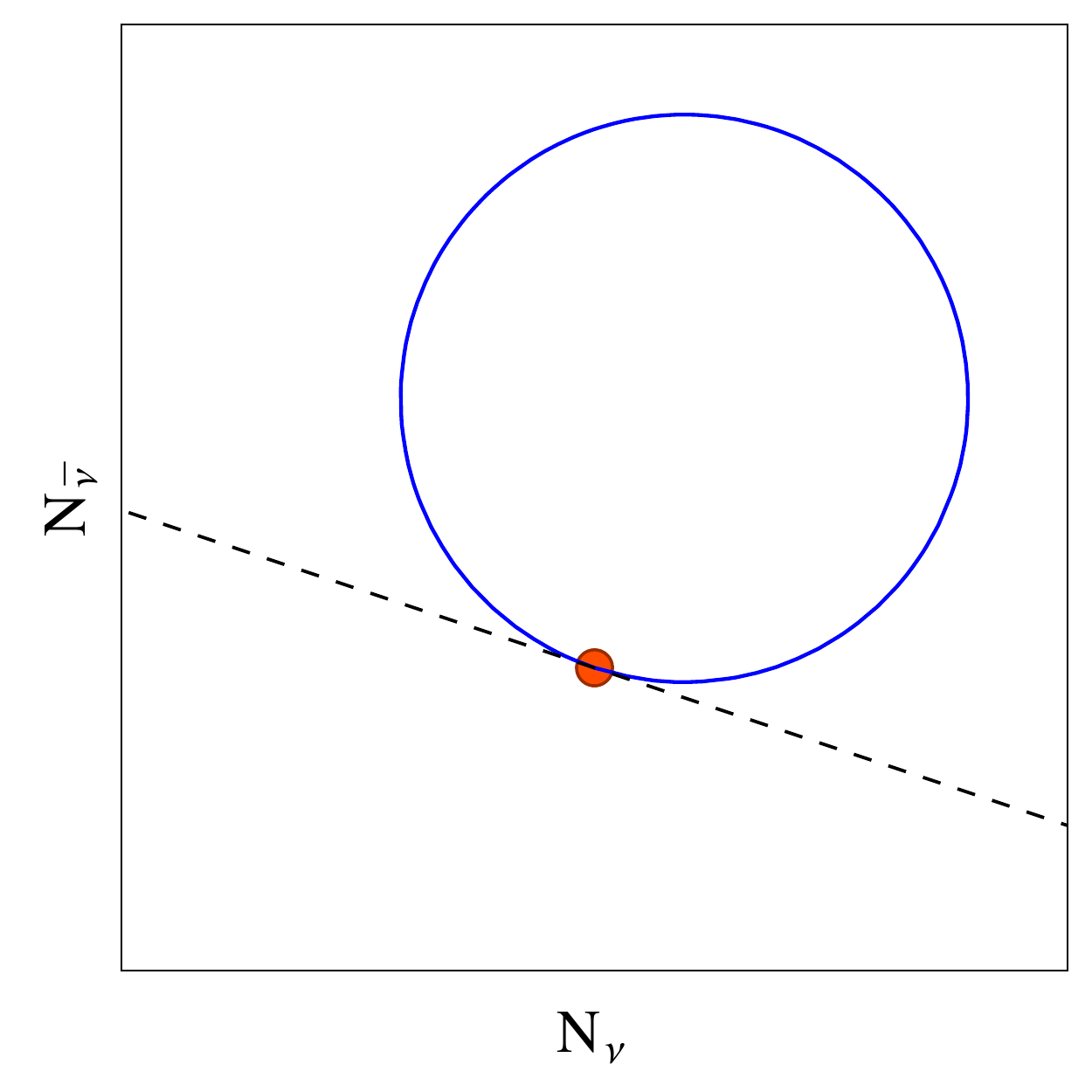}
	\includegraphics[width=0.55\textwidth]{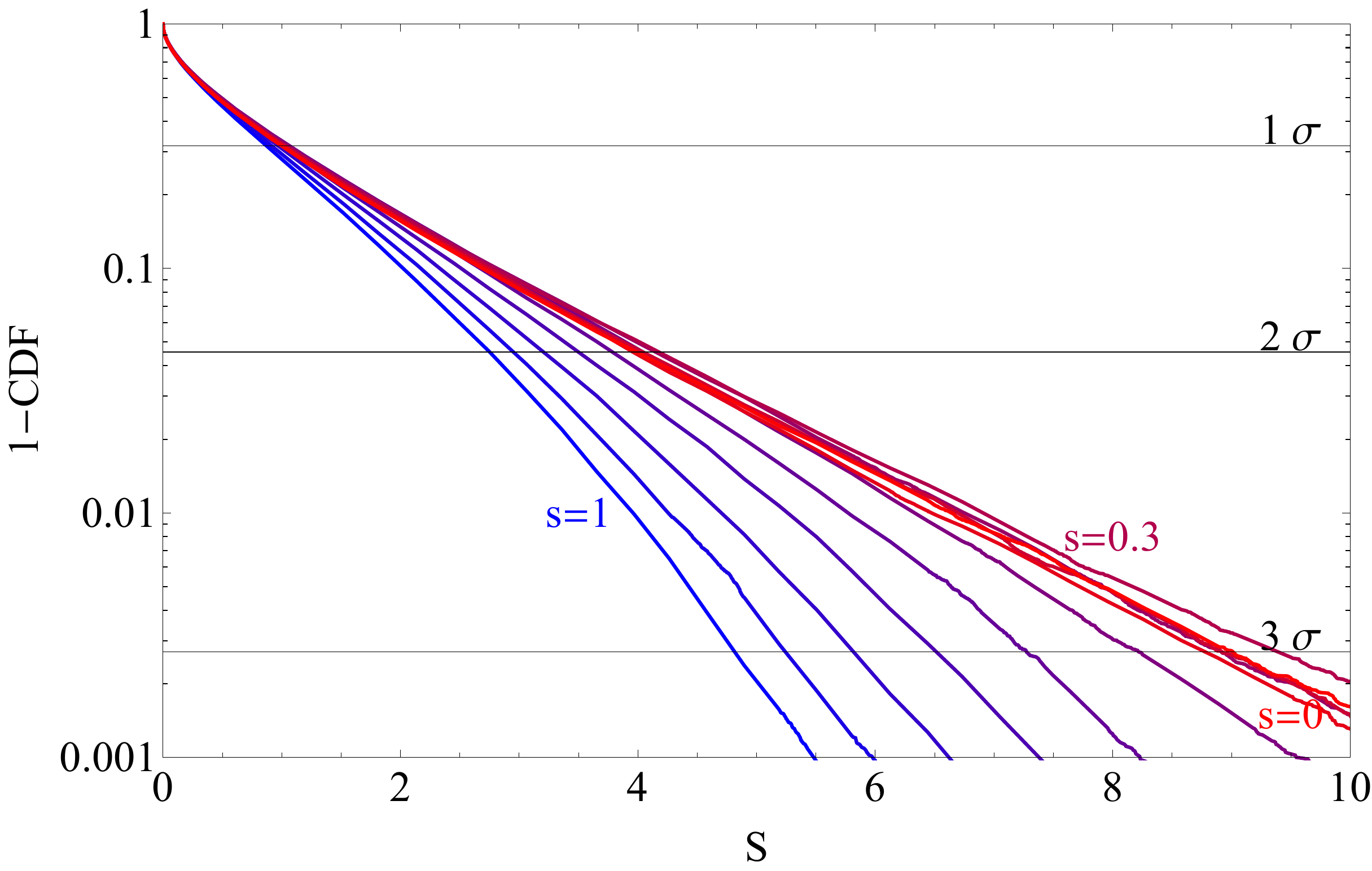}
  \caption{Left panel: toy modelization of the observable space (number of neutrino and antineutrino events) spanned by a change of $\delta$ between 0 and $2\pi$. For simplicity this has been depicted by a circle to illustrate the cyclic nature of $\delta$. The comparison with the assumptions underlying Wilk's theorem that lead to a $\chi^2$ distribution is depicted by the line tangent to the circle at the test point corresponding to $\delta=0$. Right panel: the distributions of the test statistic obtained using the model depicted in the left panel. The different lines correspond to different sizes of the standard deviation assumed for the perturbations $s$ ranging from $s = 1$ (the circle radius) to vanishing $s = 0$ in steps of 0.1. Thus lines with large $s$ would thus represent facilities with poor sensitivity to $\delta$. \label{fig:circlecheck}}
 \end{center}
\end{figure}

To understand all the features displayed by the left panel of \fig~\ref{fig:novacheck} we have used a toy model, represented in the left panel of \fig~\ref{fig:circlecheck}. Let us assume that the observables used to determine $\delta$ are only the total number of events in the apperance channels for neutrinos and antineutrinos. The expected values for the number of events would span an ellipse in observable space upon varying $\delta$ (as opposed to an infinite line, as required by the assumptions of Wilk's theorem). For simplicity we approximate this ellipse in observable space by a circle\footnote{As we will see, even with these simplifying assumptions the qualitative, and even quantitative, behaviour observed in the left panel of \fig~\ref{fig:novacheck} can be very well reproduced.}, depicted in the left panel in \fig~\ref{fig:circlecheck}. The line tangent to the circle represents how this distribution of expected number of events should look like if the premises of Wilk's theorem were satisfied. The point belonging both to the line and to the circle represents the value of $\delta = 0$, for which we wish to perform  the calibration of the test statistic. To perform this calibration, gaussian statistical fluctuations with a characteristic standard deviation $s$ (determined by the statistic and systematic errors of the experiment) must be considered around the $\delta=0$ point. In the case where Wilks' theorem holds, the test statistic for each realization would be obtained as the square of the distance of the point corresponding to the fluctuation to the $\delta = 0$ point (\ie, $\rm{min}_{\delta=0}\chi^2$ in \eq~\ref{eq:teststatistic}) minus the square of the distance of this point \emph{to the line} (\ie, $\rm{min}_{global}\chi^2$ in \eq~\ref{eq:teststatistic}), and a $\chi^2$ distribution would be obtained as a result. Here, the distances are defined relative to the standard deviation of the gaussian fluctuations. On the other hand, for the case where the variable is cyclic the test statistic would rather correspond to the square of the distance between the point corresponding to the fluctiation and the $\delta = 0$ point minus the square of the distance of this point \emph{to the circle} (again, in analogy to \eq~\ref{eq:teststatistic}). In this case, the test statistic will not necessarily be $\chi^2$-distributed anymore, as we will show below.

Facilities with poor sensitivity to $\delta$ at a given CL, \ie, facilities for which the allowed region reconstructed for $\delta$ at this CL span essentially the whole allowed range from 0 to $2 \pi$, will be characterized by statistical fluctuations (at this CL) which are of the same order or even larger than the size of the circle in the left panel of Figure~\ref{fig:circlecheck}. Indeed, if the expected fluctuations can reach the whole circle, all values of $\delta$ must necessarily be considered a good fit to the data. In this case, statistical fluctuations will often be significantly larger than the size of the circle and the distance from any fluctuation to the circle will tend to be larger than the distance of the fluctuations to the line. Hence, the value of $S$ reconstructed for this fluctuations will tend to be significantly below a $\chi^2$ distribution. We have explicitly computed the distribution of the test statistics for our toy model, for different sizes of the standard deviation $s$ relative to the circle radius depicted in the left panel in \fig~\ref{fig:circlecheck}. The obtained distributions are depicted in the right panel of \fig~\ref{fig:circlecheck} for different sizes of the standard deviation $s$ assumed for the perturbations ranging from $s = 1$ (normalizing to the circle radius) to vanishing $s = 0$ in steps of 0.1. The similarity of this figure and the right panel of \fig~\ref{fig:novacheck} is remarkable. As anticipated, the lines for which the size of the fluctuations is of the order of the circle size ($s > 0.5$) translate into a distribution falling significantly faster than a $\chi^2$ distribution with one degree of freedom. These curves can be mapped to a great accuracy to those for NO$\nu$A with factors of 1, 2 and 5 its nominal statistics. In particular, the nominal NO$\nu$A exposure seems to roughly correspond to the $s \sim 0.7$, \ie, a standard deviation for the perturbation 0.7 times the circle radius, while exposures increased by factors of 2 and 5 roughly correspond to $s \sim 0.5$ and $s \sim 0.4$ respectively. As the relative size of the gaussian perturbations with respect to the size of the circle decreases, the CDF becomes closer and closer to a $\chi^2$ distribution.

On the other hand, a large increase in the statistics for NO$\nu$A would translate in our toy model to a significant decrease in the relative size of the fluctuations with respect to the size of the circle. In this situation, perturbations can then easily fall \emph{inside} the circle. While for small size of the perturbations with respect to the circle radius there are still more points closer to the line than to the circle, the points that fall within the circle overcompensate for this fact since they are \emph{significantly} closer to it. Therefore, the average distance is shorter to the circle than to the line, and the distribution of $S$ gets shifted to larger values with respect to the $\chi^2$ distribution. This is shown by the lines obtained for $s \sim 0.3$ in \fig~\ref{fig:circlecheck}, and can be mapped to the results obtained for NO$\nu$A using 10 and 20 times higher statistics in \fig~\ref{fig:novacheck}.  

If the size of the perturbation keeps decreasing, then the distribution asymptotically approaches the $\chi^2$ distribution. This is expected, since in the limit when the perturbation is very small the system is no longer sensitive to the curvature of the outcome space and the effects coming from this should vanish. This result is shown by the red line obtained for vanishing $s$ in the right panel in \fig~\ref{fig:circlecheck}, which perfectly reproduces a $\chi^2$ distribution and is already very close also to the $s = 0.1$ line. In turn, this seems to also be the tendency shown for NO$\nu$A with 100 times increased statistics in \fig~\ref{fig:novacheck}.

Finally, we briefly comment on the role of sign degeneracies. In presence of sign degeneracies the above simplified model would have two, rather than one, ellipses (or circles). In this situation, the perturbations that take place in a direction away from the second (degenerate) ellipse are essentially unaffected by its presence. However, for perturbations in the direction of the second ellipse, the distance to this second ellipse can decrease with respect to the case in which no degeneracies are present. Naively, this could mean that the values of the test statistics $S$ would increase, and indeed this is the effect observed for T2HK in Figure~\ref{fig:cdfs}. However, the distance between the fluctuation and the CP-conserving points in the new ellipse may also decrease, which would tend to produce a decrease of the obtained values of $S$. In practice, whether the values of $S$ increase or decrease when in presence of sign degeneracies depends on the relative positions of the two ellipses in observable space as well as on the relative locations of the CP-conserving values of $\delta$ upon them. These locations are facility- (energy- and baseline-) dependent and we have found examples that change the distribution of $S$ in either direction. However, in all cases the effect was rather minor and subdominant with respect to the cyclicity of $\delta$ as described above. Moreover, the vast majority of proposed future facilities facing the measurement of $\delta$ would be either able to determine the mass hierarchy at a high CL, or far enough in the future that the mass hierarchy will most likely have been measured by some other combination of experiments. A more interesting effect, though, might be caused by the octant degeneracy allowed by the ambiguity in the value of $\theta_{23}$, as the preliminary results from~\cite{Gonzalez-Garcia:2014bfa} show. However, a detailed study of these degeneracies calls for a calibration of the test statistic as a function of both $\delta$ and $\theta_{23}$, in order to properly asses the significance of a signal in both variables simultaneously. Such a study is computationally expensive and beyond the scope of this work.

\subsection{Sensitivities reassessed}
\label{sec:sens}

\begin{figure}
 \includegraphics[width=0.7\columnwidth]{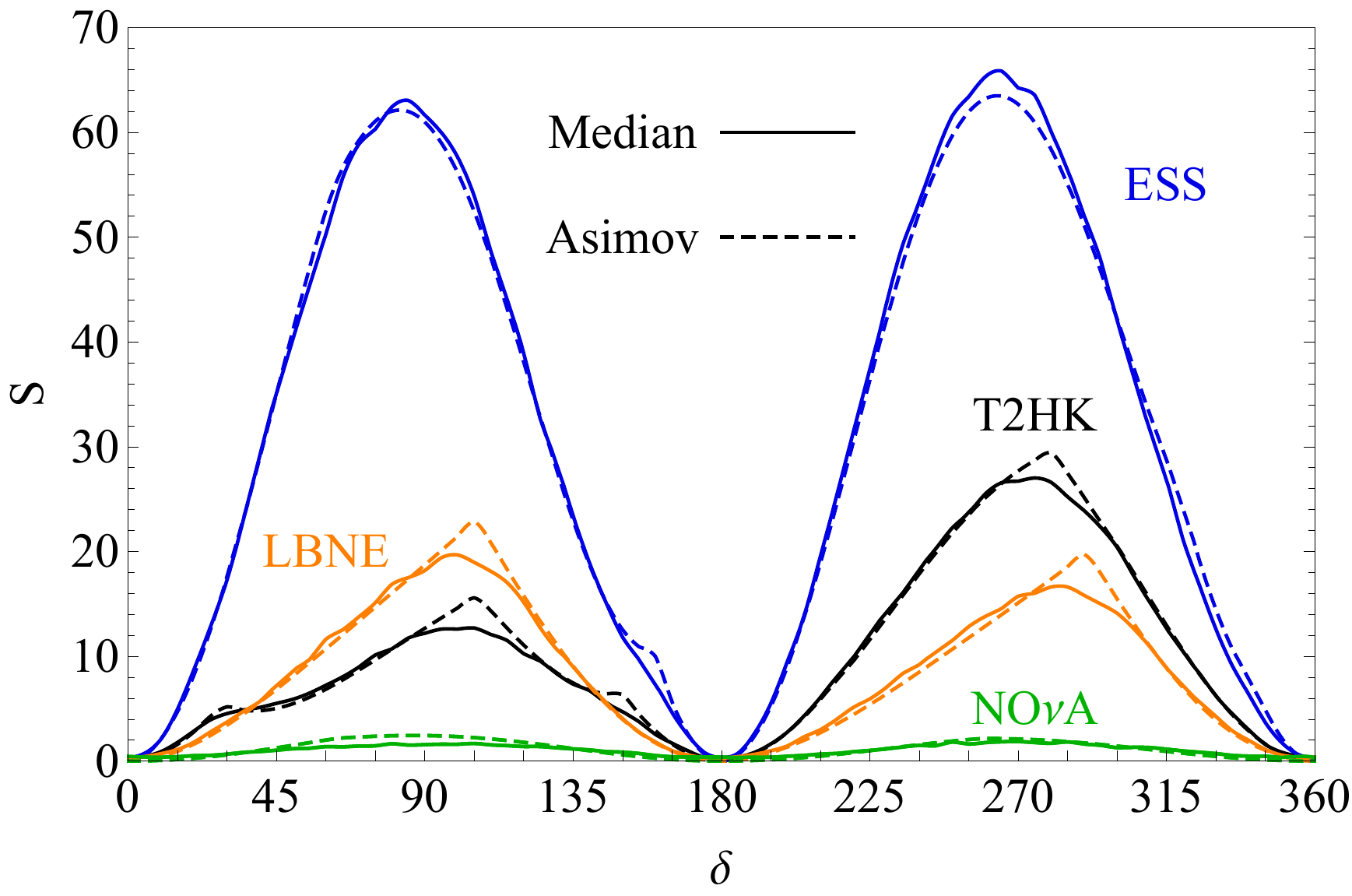}
 \begin{center}
  \caption{The test statistic $S$ for the simulated experiments, as a function of the true value of $\delta$ (in degrees). The median values obtained from Monte Carlo simulations are shown as solid lines, whereas the values corresponding to the predicted Asimov data set are shown as dashed lines.  \label{fig:median}}
 \end{center}
\end{figure}

A priori, a variation on the critical values $S_c$ with respect to the ones obtained for a $\chi^2$ distribution does not necessarily imply a change in the sensitivity of an experiment. This will depend on whether the Asimov data can be taken as a good approximation of the median result for values of $\delta \notin \{0,\pi\}$. In \fig~\ref{fig:median}, the median value of the distribution of $S$ is shown as a function of $\delta$ for all experiments under consideration, and the results are compared to the values obtained from the Asimov data. 
We find that the Asimov data set is very close to the median value of $S$ for all future experiments, while significant deviations are found for \NOvA. It thus follows that the Asimov data may be taken as a good approximation of the median test statistic when considering the more sensitive experiments, while for \NOvA\ Monte Carlo simulation would be advisable.  

\begin{figure}
 \begin{center}
  \includegraphics[width=1\columnwidth]{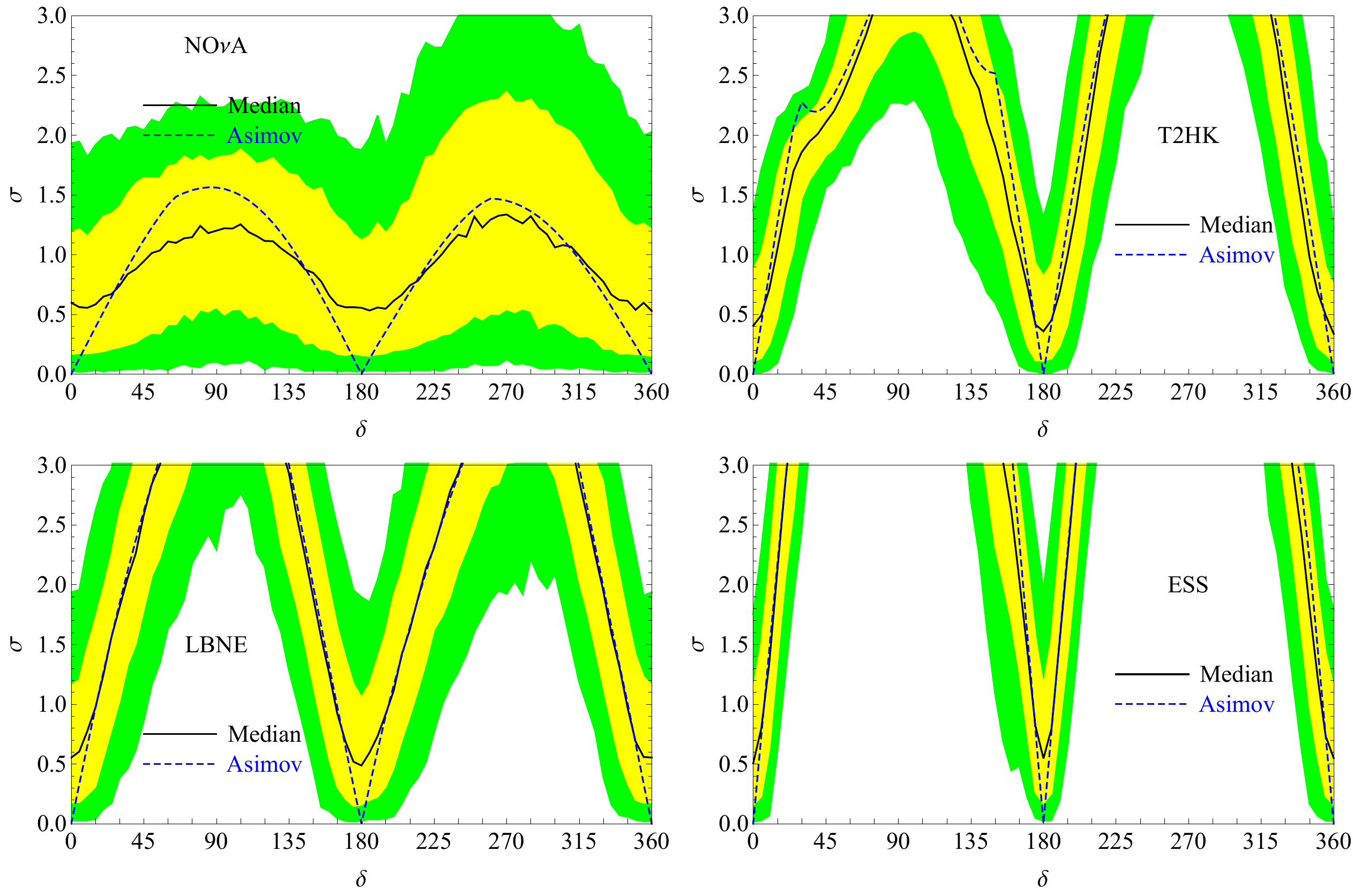}
  \caption{Predicted median sensitivity for rejecting CP conservation, as a function of the true value of $\delta$ (in degrees). Solid lines show the results using the true median and Monte Carlo calibrated distributions of $S$, while dashed lines show the results using the common approach of taking $\sqrt{S}$ for the Asimov data set. The yellow (green) bands show the regions containing 68~\% (95~\%) of the experimental realizations obtained from the Monte Carlo.  \label{fig:CPfraction}}
 \end{center}
\end{figure}

Finally, in \fig~\ref{fig:CPfraction} we compare the computed sensitivity to CP violation using a full analysis based on Monte Carlo simulation with the results obtained in the common approach (\ie, assuming the cut values of a $\chi^2$ distribution with one degree of freedom and the Asimov data set). The results obtained in 68\% and 95\% of the simulated experiments are also indicated by the yellow and green bands, respectively, to show the expected dispersion with respect to the median result.
From this figure, a reasonable agreement in the sensitivity can be observed with respect to the results obtained by simply taking the $\sqrt{S}$ for the Asimov data. The only exception is for NO$\nu$A in the region around $\delta \sim 90^\circ$ where, even if the critical value $S_c$ from \fig~\ref{fig:cdfs} is significantly lower than for a $\chi^2$ distribution, the Asimov data set considerably overestimates the median result from the Monte Carlo (see \fig~\ref{fig:median}). It is also noteworthy the sizable dispersion depicted by the 1 and 2 $\sigma$ bands for the NO$\nu$A setup that, for $\delta \sim 270^\circ$, span significances from around 0.1 to more than $3 \sigma$ in the $2 \sigma$ band. Thus, it is challenging to actually forecast the expected sensitivity for this facility, particularly since $\delta \sim 270^\circ$ happens to be the present best fit from T2K and reactor data~\cite{GonzalezGarcia:2012sz}.

We also note that the Monte Carlo results do not go to zero around the CP conserving values of $\delta$. Instead the median sensitivity is around $0.67\sigma$, corresponding to a CL of 50~\%. This should be expected as the median confidence level obtained if CP is conserved should be 50~\%. The Asimov data cannot reflect this since any fluctuations around it will increase the value of the test statistic. It is therefore not a good approximation of the median in a region close to the CP conserving values of $\delta$.

\section{Summary and outlook}

In this work, we have studied the validity of the common approach used to compute the sensitivity of future long baseline experiments to leptonic CP violation. By explicit Monte Carlo simulation we have found that the test statistic (defined in \eq~(\ref{eq:teststatistic})) is close to being $\chi^2$-distributed only when, at the corresponding confidence level which is being tested, the error bars with which $\delta$ can be reconstructed are significantly smaller than the whole $\delta$ range, so that the cyclic nature the variable is not apparent. Even in this scenario, the test statistics is slightly shifted to higher values than those obtained for a $\chi^2$ distribution. Otherwise, the distribution of the test statistic is instead significantly shifted towards lower values than naively expected by a $\chi^2$ distribution. 

We have also found that the median value of the distribution is well approximated by the Asimov data set in most cases. This results in a sensitivity which is very similar, although slightly worse, than what is typically quoted in the literature. 

In view of our results and given the present hint for $\delta \sim 270^\circ$~\cite{GonzalezGarcia:2012sz} from the combination of T2K (with relatively low statistics), and reactor data, it would be interesting to reassess its significance through a calibration of the test statistic also fully taking into account the octant degeneracy which seems to play a significant role.

\begin{acknowledgments}
We are very grateful to Thomas Schwetz for valuable discussions and encouragement. We warmly thank Peter Ballett for pointing out an inconsistency in the treatment of the systematic errors in the first version of the manuscript, and Pilar Hern\'andez for illuminating discussions. This work was supported by the G\"oran Gustafsson Foundation~[MB] and by the U.S. Department of Energy under award number \protect{DE-SC0003915}~[PC]. EFM acknowledges financial support by
the European Union through the FP7 Marie Curie Actions CIG NeuProbes (PCIG11-GA-2012-321582) and the ITN INVISIBLES (PITN-GA-2011-289442), and the Spanish
MINECO through the ``Ram\'on y Cajal'' programme (RYC2011-07710) and through the project FPA2009-09017. We also thank the Spanish MINECO (Centro de excelencia Severo Ochoa Program) under grant SEV-2012-0249 as well as the Nordita Scientific program ``News in Neutrino Physics'', where part of this work was performed.  

\end{acknowledgments}

\bibliographystyle{apsrev}

\begin{thebibliography}{32}
\expandafter\ifx\csname natexlab\endcsname\relax\def\natexlab#1{#1}\fi
\expandafter\ifx\csname bibnamefont\endcsname\relax
  \def\bibnamefont#1{#1}\fi
\expandafter\ifx\csname bibfnamefont\endcsname\relax
  \def\bibfnamefont#1{#1}\fi
\expandafter\ifx\csname citenamefont\endcsname\relax
  \def\citenamefont#1{#1}\fi
\expandafter\ifx\csname url\endcsname\relax
  \def\url#1{\texttt{#1}}\fi
\expandafter\ifx\csname urlprefix\endcsname\relax\def\urlprefix{URL }\fi
\providecommand{\bibinfo}[2]{#2}
\providecommand{\eprint}[2][]{\url{#2}}

\bibitem[{\citenamefont{Pontecorvo}(1957)}]{Pontecorvo:1957cp}
\bibinfo{author}{\bibfnamefont{B.}~\bibnamefont{Pontecorvo}},
  \bibinfo{journal}{Sov.Phys.JETP} \textbf{\bibinfo{volume}{6}},
  \bibinfo{pages}{429} (\bibinfo{year}{1957}).

\bibitem[{\citenamefont{Pontecorvo}(1958)}]{Pontecorvo:1957qd}
\bibinfo{author}{\bibfnamefont{B.}~\bibnamefont{Pontecorvo}},
  \bibinfo{journal}{Sov.Phys.JETP} \textbf{\bibinfo{volume}{7}},
  \bibinfo{pages}{172} (\bibinfo{year}{1958}).

\bibitem[{\citenamefont{Maki et~al.}(1960)\citenamefont{Maki, Nakagawa, Ohnuki,
  and Sakata}}]{Maki:1960ut}
\bibinfo{author}{\bibfnamefont{Z.}~\bibnamefont{Maki}},
  \bibinfo{author}{\bibfnamefont{M.}~\bibnamefont{Nakagawa}},
  \bibinfo{author}{\bibfnamefont{Y.}~\bibnamefont{Ohnuki}}, \bibnamefont{and}
  \bibinfo{author}{\bibfnamefont{S.}~\bibnamefont{Sakata}},
  \bibinfo{journal}{Prog.Theor.Phys.} \textbf{\bibinfo{volume}{23}},
  \bibinfo{pages}{1174} (\bibinfo{year}{1960}).

\bibitem[{\citenamefont{Maki et~al.}(1962)\citenamefont{Maki, Nakagawa, and
  Sakata}}]{Maki:1962mu}
\bibinfo{author}{\bibfnamefont{Z.}~\bibnamefont{Maki}},
  \bibinfo{author}{\bibfnamefont{M.}~\bibnamefont{Nakagawa}}, \bibnamefont{and}
  \bibinfo{author}{\bibfnamefont{S.}~\bibnamefont{Sakata}},
  \bibinfo{journal}{Prog.Theor.Phys.} \textbf{\bibinfo{volume}{28}},
  \bibinfo{pages}{870} (\bibinfo{year}{1962}).

\bibitem[{\citenamefont{Pontecorvo}(1968)}]{Pontecorvo:1967fh}
\bibinfo{author}{\bibfnamefont{B.}~\bibnamefont{Pontecorvo}},
  \bibinfo{journal}{Sov.Phys.JETP} \textbf{\bibinfo{volume}{26}},
  \bibinfo{pages}{984} (\bibinfo{year}{1968}).

\bibitem[{\citenamefont{An et~al.}(2012)}]{An:2012eh}
\bibinfo{author}{\bibfnamefont{F.}~\bibnamefont{An}} \bibnamefont{et~al.}
  (\bibinfo{collaboration}{DAYA-BAY Collaboration}),
  \bibinfo{journal}{Phys.Rev.Lett.} \textbf{\bibinfo{volume}{108}},
  \bibinfo{pages}{171803} (\bibinfo{year}{2012}), \eprint{1203.1669}.

\bibitem[{\citenamefont{Ahn et~al.}(2012)}]{Ahn:2012nd}
\bibinfo{author}{\bibfnamefont{J.}~\bibnamefont{Ahn}} \bibnamefont{et~al.}
  (\bibinfo{collaboration}{RENO collaboration}),
  \bibinfo{journal}{Phys.Rev.Lett.} \textbf{\bibinfo{volume}{108}},
  \bibinfo{pages}{191802} (\bibinfo{year}{2012}), \eprint{1204.0626}.

\bibitem[{\citenamefont{Abe et~al.}(2012)}]{Abe:2012tg}
\bibinfo{author}{\bibfnamefont{Y.}~\bibnamefont{Abe}} \bibnamefont{et~al.}
  (\bibinfo{collaboration}{Double Chooz Collaboration}),
  \bibinfo{journal}{Phys.Rev.} \textbf{\bibinfo{volume}{D86}},
  \bibinfo{pages}{052008} (\bibinfo{year}{2012}), \eprint{1207.6632}.

\bibitem[{\citenamefont{Adamson et~al.}(2011)}]{Adamson:2011qu}
\bibinfo{author}{\bibfnamefont{P.}~\bibnamefont{Adamson}} \bibnamefont{et~al.}
  (\bibinfo{collaboration}{MINOS Collaboration}),
  \bibinfo{journal}{Phys.Rev.Lett.} \textbf{\bibinfo{volume}{107}},
  \bibinfo{pages}{181802} (\bibinfo{year}{2011}), \eprint{1108.0015}.

\bibitem[{\citenamefont{Abe et~al.}(2011{\natexlab{a}})}]{Abe:2011sj}
\bibinfo{author}{\bibfnamefont{K.}~\bibnamefont{Abe}} \bibnamefont{et~al.}
  (\bibinfo{collaboration}{T2K Collaboration}),
  \bibinfo{journal}{Phys.Rev.Lett.} \textbf{\bibinfo{volume}{107}},
  \bibinfo{pages}{041801} (\bibinfo{year}{2011}{\natexlab{a}}),
  \eprint{1106.2822}.

\bibitem[{\citenamefont{Gonzalez-Garcia
  et~al.}(2012)\citenamefont{Gonzalez-Garcia, Maltoni, Salvado, and
  Schwetz}}]{GonzalezGarcia:2012sz}
\bibinfo{author}{\bibfnamefont{M.}~\bibnamefont{Gonzalez-Garcia}},
  \bibinfo{author}{\bibfnamefont{M.}~\bibnamefont{Maltoni}},
  \bibinfo{author}{\bibfnamefont{J.}~\bibnamefont{Salvado}}, \bibnamefont{and}
  \bibinfo{author}{\bibfnamefont{T.}~\bibnamefont{Schwetz}},
  \bibinfo{journal}{JHEP} \textbf{\bibinfo{volume}{1212}}, \bibinfo{pages}{123}
  (\bibinfo{year}{2012}), \eprint{1209.3023},
  \urlprefix\url{http://www.nu-fit.org/}.

\bibitem[{\citenamefont{Beringer et~al.}(2012)}]{Beringer:1900zz}
\bibinfo{author}{\bibfnamefont{J.}~\bibnamefont{Beringer}} \bibnamefont{et~al.}
  (\bibinfo{collaboration}{Particle Data Group}), \bibinfo{journal}{Phys.Rev.}
  \textbf{\bibinfo{volume}{D86}}, \bibinfo{pages}{010001}
  (\bibinfo{year}{2012}).

\bibitem[{\citenamefont{Gavela et~al.}(1994{\natexlab{a}})\citenamefont{Gavela,
  Hernandez, Orloff, and Pene}}]{Gavela:1993ts}
\bibinfo{author}{\bibfnamefont{M.}~\bibnamefont{Gavela}},
  \bibinfo{author}{\bibfnamefont{P.}~\bibnamefont{Hernandez}},
  \bibinfo{author}{\bibfnamefont{J.}~\bibnamefont{Orloff}}, \bibnamefont{and}
  \bibinfo{author}{\bibfnamefont{O.}~\bibnamefont{Pene}},
  \bibinfo{journal}{Mod.Phys.Lett.} \textbf{\bibinfo{volume}{A9}},
  \bibinfo{pages}{795} (\bibinfo{year}{1994}{\natexlab{a}}),
  \eprint{hep-ph/9312215}.

\bibitem[{\citenamefont{Gavela et~al.}(1994{\natexlab{b}})\citenamefont{Gavela,
  Hernandez, Orloff, Pene, and Quimbay}}]{Gavela:1994dt}
\bibinfo{author}{\bibfnamefont{M.}~\bibnamefont{Gavela}},
  \bibinfo{author}{\bibfnamefont{P.}~\bibnamefont{Hernandez}},
  \bibinfo{author}{\bibfnamefont{J.}~\bibnamefont{Orloff}},
  \bibinfo{author}{\bibfnamefont{O.}~\bibnamefont{Pene}}, \bibnamefont{and}
  \bibinfo{author}{\bibfnamefont{C.}~\bibnamefont{Quimbay}},
  \bibinfo{journal}{Nucl.Phys.} \textbf{\bibinfo{volume}{B430}},
  \bibinfo{pages}{382} (\bibinfo{year}{1994}{\natexlab{b}}),
  \eprint{hep-ph/9406289}.

\bibitem[{\citenamefont{Qian et~al.}(2012)\citenamefont{Qian, Tan, Wang, Ling,
  McKeown et~al.}}]{Qian:2012zn}
\bibinfo{author}{\bibfnamefont{X.}~\bibnamefont{Qian}},
  \bibinfo{author}{\bibfnamefont{A.}~\bibnamefont{Tan}},
  \bibinfo{author}{\bibfnamefont{W.}~\bibnamefont{Wang}},
  \bibinfo{author}{\bibfnamefont{J.}~\bibnamefont{Ling}},
  \bibinfo{author}{\bibfnamefont{R.}~\bibnamefont{McKeown}},
  \bibnamefont{et~al.}, \bibinfo{journal}{Phys.Rev.}
  \textbf{\bibinfo{volume}{D86}}, \bibinfo{pages}{113011}
  (\bibinfo{year}{2012}), \eprint{1210.3651}.

\bibitem[{\citenamefont{Capozzi et~al.}(2014)\citenamefont{Capozzi, Lisi, and
  Marrone}}]{Capozzi:2013psa}
\bibinfo{author}{\bibfnamefont{F.}~\bibnamefont{Capozzi}},
  \bibinfo{author}{\bibfnamefont{E.}~\bibnamefont{Lisi}}, \bibnamefont{and}
  \bibinfo{author}{\bibfnamefont{A.}~\bibnamefont{Marrone}},
  \bibinfo{journal}{Phys.Rev.} \textbf{\bibinfo{volume}{D89}},
  \bibinfo{pages}{013001} (\bibinfo{year}{2014}), \eprint{1309.1638}.

\bibitem[{\citenamefont{Blennow et~al.}(2014)\citenamefont{Blennow, Coloma,
  Huber, and Schwetz}}]{Blennow:2013oma}
\bibinfo{author}{\bibfnamefont{M.}~\bibnamefont{Blennow}},
  \bibinfo{author}{\bibfnamefont{P.}~\bibnamefont{Coloma}},
  \bibinfo{author}{\bibfnamefont{P.}~\bibnamefont{Huber}}, \bibnamefont{and}
  \bibinfo{author}{\bibfnamefont{T.}~\bibnamefont{Schwetz}},
  \bibinfo{journal}{JHEP} \textbf{\bibinfo{volume}{1403}}, \bibinfo{pages}{028}
  (\bibinfo{year}{2014}), \eprint{1311.1822}.

\bibitem[{\citenamefont{Vitells and Read}(2013)}]{Vitells:2013uza}
\bibinfo{author}{\bibfnamefont{O.}~\bibnamefont{Vitells}} \bibnamefont{and}
  \bibinfo{author}{\bibfnamefont{A.}~\bibnamefont{Read}}
  (\bibinfo{year}{2013}), \eprint{1311.4076}.

\bibitem[{\citenamefont{Feldman and Cousins}(1998)}]{Feldman:1997qc}
\bibinfo{author}{\bibfnamefont{G.~J.} \bibnamefont{Feldman}} \bibnamefont{and}
  \bibinfo{author}{\bibfnamefont{R.~D.} \bibnamefont{Cousins}},
  \bibinfo{journal}{Phys.Rev.} \textbf{\bibinfo{volume}{D57}},
  \bibinfo{pages}{3873} (\bibinfo{year}{1998}), \eprint{physics/9711021}.

\bibitem[{\citenamefont{Wilks}(1938)}]{Wilks:1938dza}
\bibinfo{author}{\bibfnamefont{S.}~\bibnamefont{Wilks}},
  \bibinfo{journal}{Annals Math.Statist.} \textbf{\bibinfo{volume}{9}},
  \bibinfo{pages}{60} (\bibinfo{year}{1938}).

\bibitem[{\citenamefont{Cowan et~al.}(2011)\citenamefont{Cowan, Cranmer, Gross,
  and Vitells}}]{Cowan:2010js}
\bibinfo{author}{\bibfnamefont{G.}~\bibnamefont{Cowan}},
  \bibinfo{author}{\bibfnamefont{K.}~\bibnamefont{Cranmer}},
  \bibinfo{author}{\bibfnamefont{E.}~\bibnamefont{Gross}}, \bibnamefont{and}
  \bibinfo{author}{\bibfnamefont{O.}~\bibnamefont{Vitells}},
  \bibinfo{journal}{Eur.Phys.J.} \textbf{\bibinfo{volume}{C71}},
  \bibinfo{pages}{1554} (\bibinfo{year}{2011}), \eprint{1007.1727}.

\bibitem[{\citenamefont{Schwetz}(2007)}]{Schwetz:2006md}
\bibinfo{author}{\bibfnamefont{T.}~\bibnamefont{Schwetz}},
  \bibinfo{journal}{Phys.Lett.} \textbf{\bibinfo{volume}{B648}},
  \bibinfo{pages}{54} (\bibinfo{year}{2007}), \eprint{hep-ph/0612223}.

\bibitem[{\citenamefont{Abe et~al.}(2011{\natexlab{b}})\citenamefont{Abe, Abe,
  Aihara, Fukuda, Hayato et~al.}}]{Abe:2011ts}
\bibinfo{author}{\bibfnamefont{K.}~\bibnamefont{Abe}},
  \bibinfo{author}{\bibfnamefont{T.}~\bibnamefont{Abe}},
  \bibinfo{author}{\bibfnamefont{H.}~\bibnamefont{Aihara}},
  \bibinfo{author}{\bibfnamefont{Y.}~\bibnamefont{Fukuda}},
  \bibinfo{author}{\bibfnamefont{Y.}~\bibnamefont{Hayato}},
  \bibnamefont{et~al.} (\bibinfo{year}{2011}{\natexlab{b}}),
  \eprint{1109.3262}.

\bibitem[{\citenamefont{Coloma et~al.}(2014)\citenamefont{Coloma, Minakata, and
  Parke}}]{Coloma:2014kca}
\bibinfo{author}{\bibfnamefont{P.}~\bibnamefont{Coloma}},
  \bibinfo{author}{\bibfnamefont{H.}~\bibnamefont{Minakata}}, \bibnamefont{and}
  \bibinfo{author}{\bibfnamefont{S.~J.} \bibnamefont{Parke}},
  \bibinfo{journal}{Phys.Rev.} \textbf{\bibinfo{volume}{D90}},
  \bibinfo{pages}{093003} (\bibinfo{year}{2014}), \eprint{1406.2551}.

\bibitem[{\citenamefont{Gonzalez-Garcia
  et~al.}(2014)\citenamefont{Gonzalez-Garcia, Maltoni, and
  Schwetz}}]{Gonzalez-Garcia:2014bfa}
\bibinfo{author}{\bibfnamefont{M.}~\bibnamefont{Gonzalez-Garcia}},
  \bibinfo{author}{\bibfnamefont{M.}~\bibnamefont{Maltoni}}, \bibnamefont{and}
  \bibinfo{author}{\bibfnamefont{T.}~\bibnamefont{Schwetz}},
  \bibinfo{journal}{JHEP} \textbf{\bibinfo{volume}{1411}}, \bibinfo{pages}{052}
  (\bibinfo{year}{2014}), \eprint{1409.5439}.

\bibitem[{\citenamefont{Adams et~al.}(2013)}]{Adams:2013qkq}
\bibinfo{author}{\bibfnamefont{C.}~\bibnamefont{Adams}} \bibnamefont{et~al.}
  (\bibinfo{collaboration}{LBNE Collaboration}) (\bibinfo{year}{2013}),
  \eprint{1307.7335}.

\bibitem[{\citenamefont{Baussan et~al.}(2014)}]{Baussan:2013zcy}
\bibinfo{author}{\bibfnamefont{E.}~\bibnamefont{Baussan}} \bibnamefont{et~al.}
  (\bibinfo{collaboration}{ESSnuSB Collaboration}), \bibinfo{journal}{Nuclear
  Physics B} \textbf{\bibinfo{volume}{885}}, \bibinfo{pages}{127 }
  (\bibinfo{year}{2014}), \eprint{1309.7022}.

\bibitem[{\citenamefont{Ayres et~al.}(2004)}]{Ayres:2004js}
\bibinfo{author}{\bibfnamefont{D.}~\bibnamefont{Ayres}} \bibnamefont{et~al.}
  (\bibinfo{collaboration}{NOvA Collaboration}) (\bibinfo{year}{2004}),
  \eprint{hep-ex/0503053}.

\bibitem[{\citenamefont{Huber et~al.}(2005)\citenamefont{Huber, Lindner, and
  Winter}}]{Huber:2004ka}
\bibinfo{author}{\bibfnamefont{P.}~\bibnamefont{Huber}},
  \bibinfo{author}{\bibfnamefont{M.}~\bibnamefont{Lindner}}, \bibnamefont{and}
  \bibinfo{author}{\bibfnamefont{W.}~\bibnamefont{Winter}},
  \bibinfo{journal}{Comput.Phys.Commun.} \textbf{\bibinfo{volume}{167}},
  \bibinfo{pages}{195} (\bibinfo{year}{2005}), \eprint{hep-ph/0407333}.

\bibitem[{\citenamefont{Huber et~al.}(2007)\citenamefont{Huber, Kopp, Lindner,
  Rolinec, and Winter}}]{Huber:2007ji}
\bibinfo{author}{\bibfnamefont{P.}~\bibnamefont{Huber}},
  \bibinfo{author}{\bibfnamefont{J.}~\bibnamefont{Kopp}},
  \bibinfo{author}{\bibfnamefont{M.}~\bibnamefont{Lindner}},
  \bibinfo{author}{\bibfnamefont{M.}~\bibnamefont{Rolinec}}, \bibnamefont{and}
  \bibinfo{author}{\bibfnamefont{W.}~\bibnamefont{Winter}},
  \bibinfo{journal}{Comput.Phys.Commun.} \textbf{\bibinfo{volume}{177}},
  \bibinfo{pages}{432} (\bibinfo{year}{2007}), \eprint{hep-ph/0701187}.

\bibitem[{\citenamefont{Coloma et~al.}(2013)\citenamefont{Coloma, Huber, Kopp,
  and Winter}}]{Coloma:2012ji}
\bibinfo{author}{\bibfnamefont{P.}~\bibnamefont{Coloma}},
  \bibinfo{author}{\bibfnamefont{P.}~\bibnamefont{Huber}},
  \bibinfo{author}{\bibfnamefont{J.}~\bibnamefont{Kopp}}, \bibnamefont{and}
  \bibinfo{author}{\bibfnamefont{W.}~\bibnamefont{Winter}},
  \bibinfo{journal}{Phys.Rev.} \textbf{\bibinfo{volume}{D87}},
  \bibinfo{pages}{033004} (\bibinfo{year}{2013}), \eprint{1209.5973}.

\bibitem[{\citenamefont{Blennow and Fernandez-Martinez}(2010)}]{Blennow:2009pk}
\bibinfo{author}{\bibfnamefont{M.}~\bibnamefont{Blennow}} \bibnamefont{and}
  \bibinfo{author}{\bibfnamefont{E.}~\bibnamefont{Fernandez-Martinez}},
  \bibinfo{journal}{Comput.Phys.Commun.} \textbf{\bibinfo{volume}{181}},
  \bibinfo{pages}{227} (\bibinfo{year}{2010}), \eprint{0903.3985}.

\end{thebibliography}

\end{document}